\documentclass[fleqn,usenatbib]{mnras}

\usepackage[dvipsnames]{xcolor}

\usepackage{graphicx}	

\def\ae{AE~Aqr}

\newcommand{\caps}[1]{{\scshape{#1}}}

\def\skytel{Sky \& Tel.}
\def\acta{Acta Astron.}
\def\apj{ApJ}
\def\aj{AJ}
\def\mnras{MNRAS}
\def\pasp{PASP}
\def\aap{A\&A}

\title[High-Dispersion Spectroscopy of AE Aqr II]{High-Dispersion Spectroscopy of AE Aqr -- II: evidence of material orbiting the primary star}

\author[Ram\'irez and Echevarr\'ia]{S. H. Ram\'irez $^{1}$\thanks{e-mail: 
	sergio.ramirez@ciencias.unam.mx} and
    J. Echevarr\'ia$^{1}$\\
$^{1}$ Instituto de Astronom\'ia, Universidad Nacional Aut\'onoma de M\'exico, Apartado Postal 70-264, 
~~Ciudad Universitaria, M\'exico D.F., C.P. 04510, M\'exico.
}

\date{Accepted XXX. Received YYY; in original form ZZZ}

\pubyear{2020}

\begin{document}
\label{firstpage}
\pagerange{\pageref{firstpage}--\pageref{lastpage}}
\maketitle
%
\begin{abstract}
We present a second paper of the analyses of high-dispersion spectroscopic observations of the magnetic cataclysmic variable AE Aquarii. We focus our efforts on the study of the emission lines and their radial velocities. We detect a sinusoidal behaviour, in several of the observing runs, with variable amplitudes.  Of those runs presented, the velocity curve of 2000 August shows less instability in the emission material. In this case we obtain $K_1~=~114 ~\pm ~8$~kms$^{-1}$, which we take as our best value for the radial velocity of the primary. This result is consistent within $2\sigma$ with previously published values obtained using indirect methods.  We interpret this consistency as observational evidence of material orbiting the rapidly-rotating primary star. We present a Doppler Tomography study, which shows that the $H\alpha$ emission is primarily concentrated within a blob in the lower left quadrant; a structure similar to that predicted by the propeller model. However, for 2000 August, we find the emission centred around the position of the white dwarf, which supports the possibility of the $K_1$ value of this run of being a valid approximation of the orbital motion of the white dwarf.


\end{abstract}




\section{Introduction}
Cataclysmic variables (CVs) are interacting binaries, that consist of a white dwarf (WD) primary star and a  late--type main sequence star (secondary), that transfers matter to the primary via Roche Lobe overflow  \citep[see][and references therein]{warner:1995}. \par
In the classical model, established by \citet{sma71} and \citet{wan71}, the material transferred from the secondary flows through the inner Lagrangian point and orbits around the primary forming an accretion disc. 
However, if the magnetic field of the WD is very strong ($\geq 10$~MG), it inhibits the formation of a disc and the material is channeled by the field lines towards the magnetic poles of the WD, as is the case in the so called polar systems \citep[see][and references therein]{kafka:2005}. Furthermore the magnetic field lines can lock the rotation of the compact star with the orbital period \citep[see][]{Cropper:1990}. Alternately, if the magnetic field is not strong enough (0.1-10 MG), a partial external disc determined by the Alfven radius is formed, where the internal part of the disc is threaded by the field, and caused to flow along the field lines. In this case the strength of the field is not sufficient to synchronise the rotation of the WD with the orbital period. These systems are known as intermediate polars \citep[see][and references therein]{patterson-1994}. \par
Given that CVs are spectroscopic binaries, the study of the radial velocities of their components is paramount to obtain the orbital parameters. Moreover, by means of a method called Doppler Tomography, developed by \citet{Marsh:1988},  the Doppler shifts of the emission lines are analysed as a function of the orbital phase, to create a two-dimensional map of the binary system in velocity space. \par
AE Aqr is a nova-like CV whose components are a rapidly rotating magnetic WD, with a 33~s spin period \citep{Patterson-1979}; and a late-type companion with a spectral type of K0-K4 \citep[][hereinafter Paper I]{echevarria-2008}. The binary has an unusually long orbital period of 9.88~h \citep{Walker-1965}. \par
The system was first characterised as an intermediate polar by \citet{Patterson-1979}. The moderate intensity of the WD's magnetic field \citep{Cropper:1986} should permit the partial formation of a disc.  However, spectroscopic observations do not show a double-peaked structure in the Balmer emission lines \citep[e.g.][]{Robinson:1991,welsh:1998}; a structure characteristic in the line profiles of discs in systems of high inclination \citep[see][]{Horne_1986}. Furthermore, no signature of a disc was found in the Doppler tomograms put forward by \citet{Wynn_1997} and \citet{welsh:1998}. These authors, with proper caution, interpreted this as possible evidence that the material transferred from the secondary was being ejected by the rapidly-rotating magnetic field of the WD, which acted as a propeller at the time of their observations. Indeed, numerical magnetohydrodynamics simulations \citep{Isakova:2016,blinova-2019} predict that, when in a propeller regime, an oscillating and rather unstable disc of a transient nature, is able to form around the WD of \ae, which would be difficult to detect in Doppler tomograms. 
So as was succinctly put by \citet{Wynn_1997}: \textit{AE Aqr is likely to alternate between phases of disc accretion, in which the white dwarf spins up, and propeller states in
which it spins down. } \par
In this article, we use the same UCLES and Echelle observations published in Paper I, who deferred the analysis of the emission lines for a later study. In particular, we use the observations of 1991 Aug and 2000 Aug (See Section 2 of Paper I). We also analyse an unpublished third set of observations from 2008 Aug. We will briefly describe all three observing runs in Section~\ref{sec:obs}. We present a radial velocity analysis in Section~\ref{sec:radvel}, followed by a Doppler Tomography study in Section~\ref{sec:tomo}. We close this paper with a discussion of the results in Section~\ref{discussion} and our conclusions in Section~\ref{conclusions}.

\section{Observations}
\label{sec:obs}

Spectroscopic observations of AE Aqr were made using the
Anglo-Australian Telescope (AAT) and the University College London
Echelle Spectrograph (UCLES) at the coud\'e focus, on 1991, August 2
and 3. A total of 102 spectra of 360~s exposure each, were obtained, with a typical signal to noise ratio of 10 (around  $\lambda$4500~\AA). We used a
31.6 lines~mm$^{-1}$ grating and a Blue Thomson 1024~x~1024
CCD with the 700~mm camera. The spectral region of $\lambda$4000~\AA~
to $\lambda$5100~\AA~ was covered,  with a spectral resolution of about
5.4 kms$^{-1}$. Almost two complete orbital periods were covered.
Several late-type standard stars were also observed.\par
Two further runs were carried out at the Observatorio Astron\'omico
Nacional at San Pedro M\'artir (SPM) using the 2.1m Telescope and
the Echelle Spectrograph. \par
In the second set of
observations, made during the nights of 2000 August 17 and 19,
81 spectra of 600~s exposure were obtained, using a $15 \mu$ Thomson 2048~x~2048 detector and a 300 lines~mm$^{-1}$ echellette grating to cover a spectral range from $\lambda$3700~\AA~to $\lambda$7700~\AA, with a typical signal to noise ratio of 42 (around $\lambda \, 5450$ \AA).

The third run of observations were made during the nights of 2008 August 14 and 16-20. A total of 171 spectra of 600 s exposure were obtained. A 15$\mu$m SITe3 1024 x 1024 detector was employed, covering a spectral interval from $\lambda3970$ to $\lambda6650$ \AA, with a typical signal to noise ratio of 30 (around $\lambda 5450$ \AA).

\section{Radial velocity analysis of the primary star}
\label{sec:radvel}

We measured the radial movement of the wings of the H$\alpha$ and H$\beta$ lines, using two Gaussian functions with a fixed width and separation as described by \citet{Shafter:1986}. We fit the radial velocities measured from the full set of spectra to a circular orbit: 
\begin{equation}
V(t) = \gamma + K \sin\left(2\pi\frac{t - t_0}{P_{orb}}\right),
\label{radvel}
\end{equation}
where $\gamma$ is the systemic velocity, $K$ the semi-amplitude, $t_0$ the time of inferior conjunction of the donor, and $P_{orb}$ is the orbital period. We employed $\chi^2$ as our goodness-of-fit parameter.  Following the methodology carried out by \citet{horne_etal:1986}, the orbital period was fixed in Equation \ref{radvel}, and therefore only the other three parameters were calculated. Since the orbital period has been well established in Paper I, we use the published  value of 0.4116554800(2)~days.

\begin{figure}

	\includegraphics[angle=0,height=5cm,width=1.0\columnwidth]{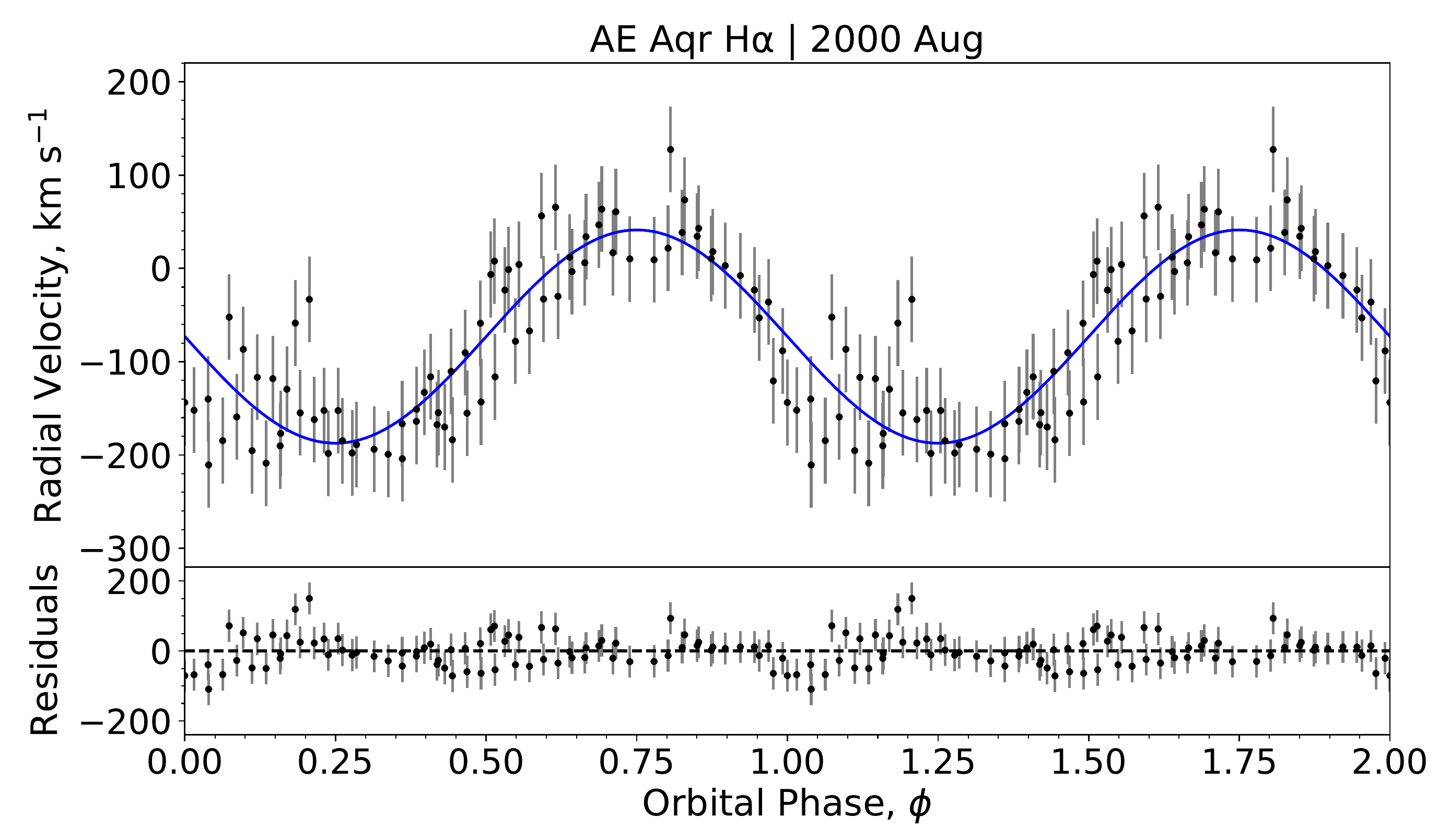}
    \caption{Radial velocity curve of $H\alpha$ for the 2000 Aug observation run. }

    \label{fig:radvel-00}
\end{figure}

\begin{figure}

	\includegraphics[angle=0,height=5cm,width=1.0\columnwidth]{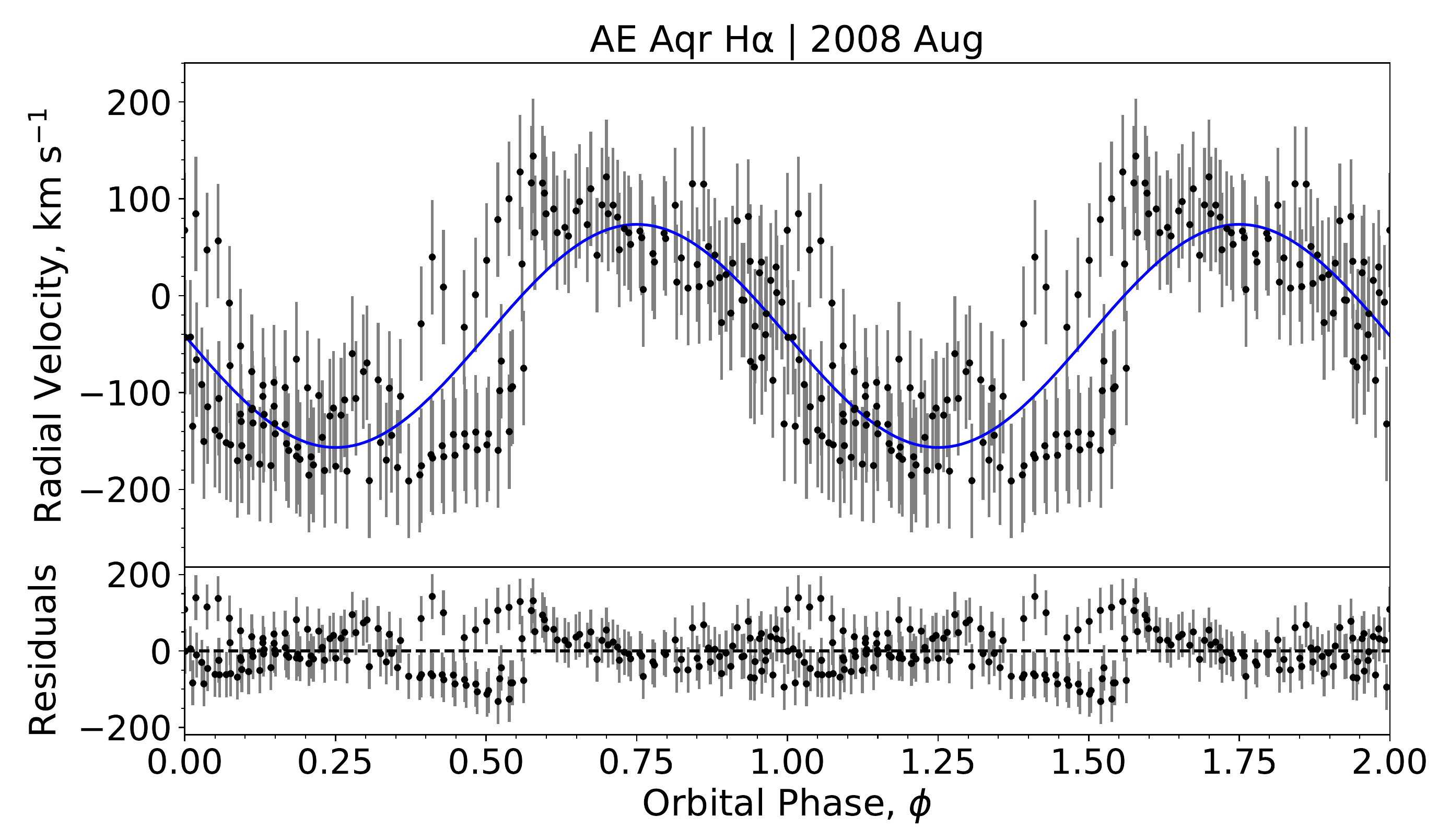}
    \caption{Radial velocity curve of $H\alpha$ for the 2008 Aug observation run.}

    \label{fig:radvel-08}
\end{figure}

\begin{figure}

	\includegraphics[angle=0,height=5cm,width=1.0\columnwidth]{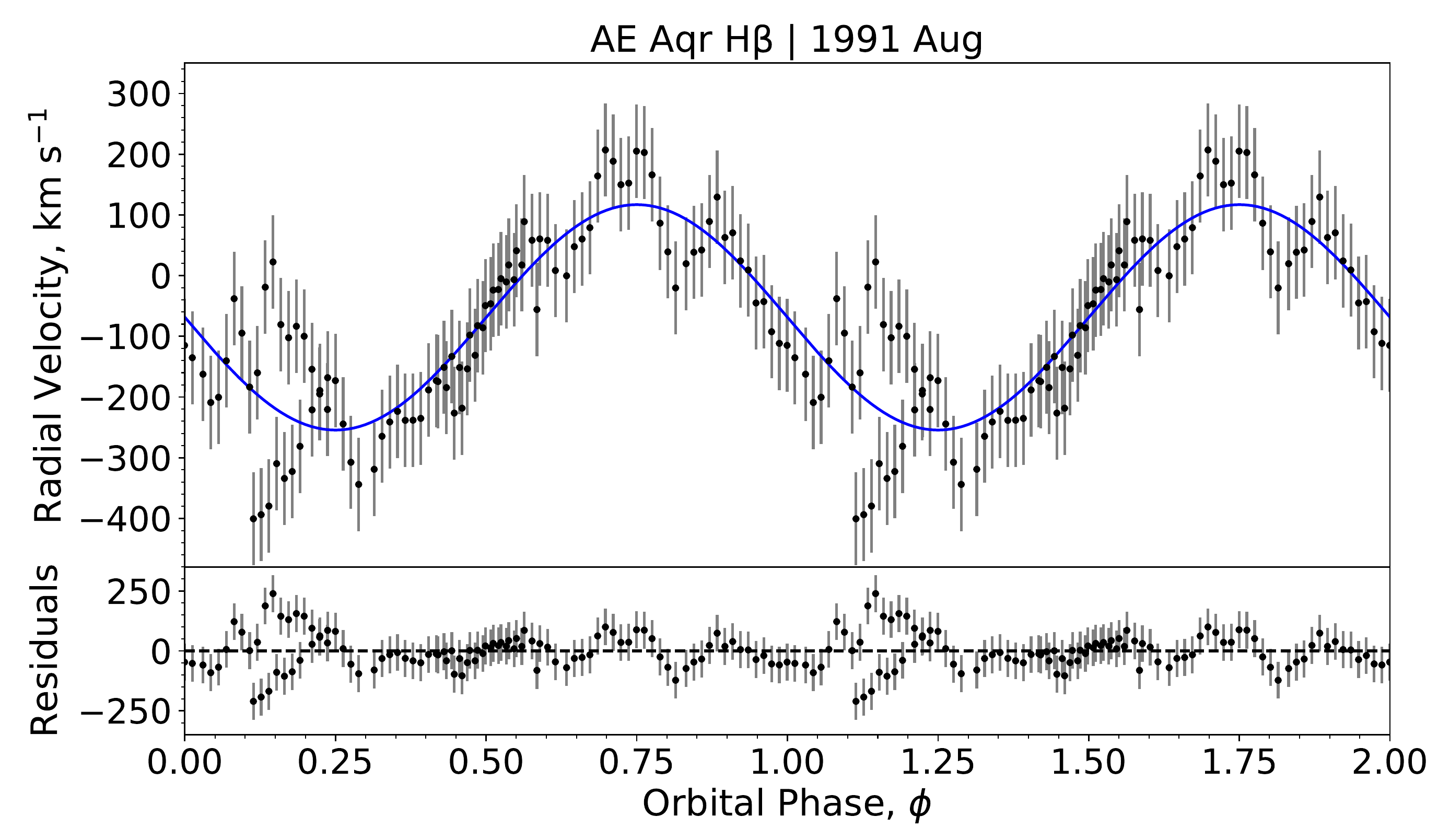}
    \caption{Radial velocity curve of $H\beta$ for the 1991 Aug observation run.  }

    \label{fig:radvel-91}
\end{figure}

\subsection{Orbital parameters}
\label{orbparam}
To measure the radial velocity of the emission line wings, we have used the {\sc convrv} task within the {\sc rvsao} package in {\sc iraf}\footnote{ IRAF is distributed by the National Optical Astronomy Observatories, which are operated by the Association of Universities for Research in Astronomy, Inc., under cooperative agreement with the National Science Foundation.} written by J. Thorstensen (2008, private communication), which convolves the line profile with an antisymmetric function and interprets the root of the convolution as the midpoint of the line profile, as per the algorithm described by \citet{schneider:1980}. 
In particular we  implemented the standard double-Gaussian ({\sc gau2}) option as our convolution function \citep[][]{Shafter:1986}, which traces the radial velocity of the profile wings. This method requires us to specify two parameters: a Gaussian width and the separation between the centres of the antisymmetric pair of Gaussians.
\par
Ideally a diagnostic diagram  is performed to probe for the optimal values of the parameters of the convolution function \citep[][]{shafter:1983,Shafter:1986}. However, given that the H$\alpha$ and H$\beta$ emission lines of AE Aqr exhibit a very unstable behaviour, we were unable to perform an enough number of stable convolutions to make a diagnostic diagram. Therefore we had to visually inspect the convolutions that yielded a solution, and chose the parameters that produced the most stable sinusoidal pattern. These parameters are listed in Table~\ref{conv-par}, while the radial velocity curves produced with these parameters are depicted in Figures \ref{fig:radvel-00}, \ref{fig:radvel-08}, and \ref{fig:radvel-91}. Note that in some phase intervals the radial velocities show considerable fluctuations and therefore the sinusoidal fit is not a simple one.
\par
We are aware that this is not the optimal approach to find the best $K1$, $\gamma$, and $HJD_0$ values, but it will serve the purpose of  detecting a sinusoidal modulation that will aid us in making a good approximation for the semi-amplitude $K_1$ by fitting the data to a circular orbit (see Equation~\ref{radvel}).
\par
\subsubsection{H$\alpha$ emission line}
\label{sec:radvel-ha}
For the observations of Aug 2000, we used a Gaussian width of 35 pixels and a separation of 122 pixels. The radial velocity curve of this run is displayed in Figure \ref{fig:radvel-00}. The orbital fit (blue curve) yielded a semiamplitude value of $K_1=114\pm8 ~kms^{-1}$.  \par

The Aug 2008 radial velocities, shown in Figure~\ref{fig:radvel-08}, were produced with a Gaussian width of 14 pixels and a separation of 121 pixels, yielding $K_1=115 \pm 6 ~kms^{-1}$, which is consistent with the value of Aug 2000. However, this run shows more variability, being especially erratic throughout the phase interval $\sim0.25-0.50$. The orbital parameters of H$\alpha$ are shown in Table \ref{orbpar1}.\par
Because the radial velocity data of Aug 2000 is the most stable of the three runs analysed, we adopt $K_1=114\pm8~kms^{-1}$ as our best approximation to the orbital motion of the white dwarf. This value is consistent within $2\sigma$ to that obtained in Paper I, wherein the rotational velocity of the secondary star was measured to indirectly derive $K_1=101\pm3~kms^{-1}$. The same consistency holds with the value derived from the pulse timing data by \citet{eracleous:1994}, of $K_1=102\pm2~kms^{-1}$.\par Taking our direct measurement of the value $K_1~=~114~\pm~8~kms^{-1}$; the semiamplitude of the secondary  $K_2~=~168~\pm~1~kms^{-1}$, obtained in Paper I; and using the fixed value of the orbital period of $P=0.4116554800(2)$, we  calculate:

\begin{equation}
M_1sin^3i=\frac{PK_2(K_1+K_2)^2}{2\pi G}=0.58 \pm 0.03 M_{\odot}
\label{eq:m1}
\end{equation}
and,
\begin{equation}
M_2sin^3i=\frac{PK_1(K_1+K_2)^2}{2\pi G}=0.39 \pm 0.05 M_{\odot}
\label{eq:m2}
\end{equation}
Assuming an inclination of $i=70^{\circ}\pm3$ yields a value of the mass of the white dwarf of $M_1=0.69\pm0.06~M_{\odot}$; while for the secondary it results in a mass of $M_2=0.47\pm0.07~M_{\odot}$. These values are consistent within the errors with those obtained in Paper I. However, the indirect derivation of $K_1$ is still more reliable than our direct measurement. Therefore, we  will adopt the values of the mass parameters published in Paper I, of $M_1=0.63\pm0.05~M_{\odot}$ and $M_2=0.37\pm0.04~M_{\odot}$, to produce the Doppler tomograms in Section~\ref{sec:tomo}.

\subsubsection{H$\beta$ emission line}
\label{sec:radvel-hb}
Variations in the line profile of H$\beta$ for the 1991 observations rendered a double-Gaussian whose parameters were difficult to find. A Gaussian width of 250 pixels and a separation of 507 pixels were used. This run presents considerable fluctuations in the phase intervals of $\sim0.00-0.25$ and $\sim0.60-0.85$, as shown in Figure~\ref{fig:radvel-91}, where the semi-amplitude yielded a high value of $K_1=185\pm11~kms^{-1}$. The orbital parameters of this run are exhibited in Table~\ref{orbpar2}. The semi-amplitude value of $H\beta$ is inconsistent from those obtained above for $H\alpha$; we address a possible cause for this deviation in Section~\ref{subsec:tomo-hb}.\par 


\begin{table}
\centering
\caption{Parameters of the used convolution functions (Double Gaussian). See text.} 
\label{conv-par}
\begin{tabular}{lccc}
\hline
Conv. Func.          &   Aug 2000  &     Aug 2008  &Aug 1991    \\
Parameter        &   H$\alpha$   &   H$\alpha$ & H$\beta$            \\
\hline
  Width      & 35   &  14  &250      \\
  (Pixels) &        &     &         \\
  Separation & 122      &  121& 507       \\
  (Pixels) &         &     &           \\
  
\hline

\end{tabular}
\end{table}

\begin{table}
\centering
\caption{Orbital Parameters obtained from the H$\alpha$ line.} 
\label{orbpar1}
\begin{tabular}{lccc}
\hline
Orbital          &   Aug 2000  &     Aug 2008      \\
Parameter        &   H$\alpha$   &   H$\alpha$             \\
\hline
  $\gamma$       & -73$\pm 5$    &  -41$\pm 4$         \\
  (km\,s$^{-1}$) &            &             \\
  $K_1$  & 114 $\pm 8$      &  115 $\pm 6$        \\
  (km\,s$^{-1}$) &         &                \\
  $HJD_{0}$  & 0.865 $\pm 0.004$  & 0.967 $\pm 0.003$  \\
  (2439030+)    &                  &       \\
  $P_{orb}$*      &  0.4116554800(2) &  0.4116554800(2)   \\
   (days)        &          &           \\
\hline
*Fixed
\end{tabular}
\end{table}

\begin{table}
\centering
\caption{Orbital Parameters obtained from the H$\beta$ line.} 
\label{orbpar2}
\begin{tabular}{lcc}
\hline
Orbital          &   Aug 1991    \\
Parameter        &   H$\beta$              \\
\hline
  $\gamma$       & -68$\pm 7$          \\
  (km\,s$^{-1}$) &                \\
  $K_1$  & 185 $\pm 11$         \\
  (km\,s$^{-1}$) &             \\
  $HJD_{0}$  & 0.840  $\pm 0.003$  \\
  (2439030+)    &                  \\
  $P_{orb}$*      &  0.4116554800(2) \\
   (days)        &            \\
\hline
*Fixed
\end{tabular}
\end{table}

\section{Doppler tomography}
\label{sec:tomo}
Doppler Tomography is a  spectroscopy technique that uses the phase resolved emission line profiles to map the accretion flow in velocity space.
 A detailed formulation of this technique can be found in \citet{Marsh:1988}.\par
 In Section \ref{subsec:tomo-ha}, we describe the Tomography of the H$\alpha$ lines from Aug 2000 and Aug 2008, and show the resulting images in Figure~\ref{fig:dopmap-ha}. To produce said tomograms, we used a newly developed \caps{pydoppler}\footnote{Available at \url{https://github.com/Alymantara/pydoppler}} python code. This code uses the original {\sc fortran} programs, developed by \citet{Spruit:1998} for an {\sc idl} enviroment. As explained in Section \ref{sec:radvel-ha}, we adopted the mass parameters of $M_1=0.63\pm0.05~M_{\odot}$ and $M_2=0.37\pm0.04~M_{\odot}$, calculated in Paper I, to produce the Tomography images.\par
 We were not able to construct the $H\beta$ tomogram from the observing run of Aug 1991, as the variations of this emission line throughout the orbital period did not allow the convergence of a reliable image. Instead we present the phase resolved profile of the emission line in Figure~\ref{fig:profiles-hb}. This variations are described in detail in Section~\ref{subsec:tomo-hb}.
 
\subsection{H{\ensuremath{\alpha}} Tomography}
 \label{subsec:tomo-ha}
\begin{figure*}
	\includegraphics[height=6cm,width=1.0\columnwidth]{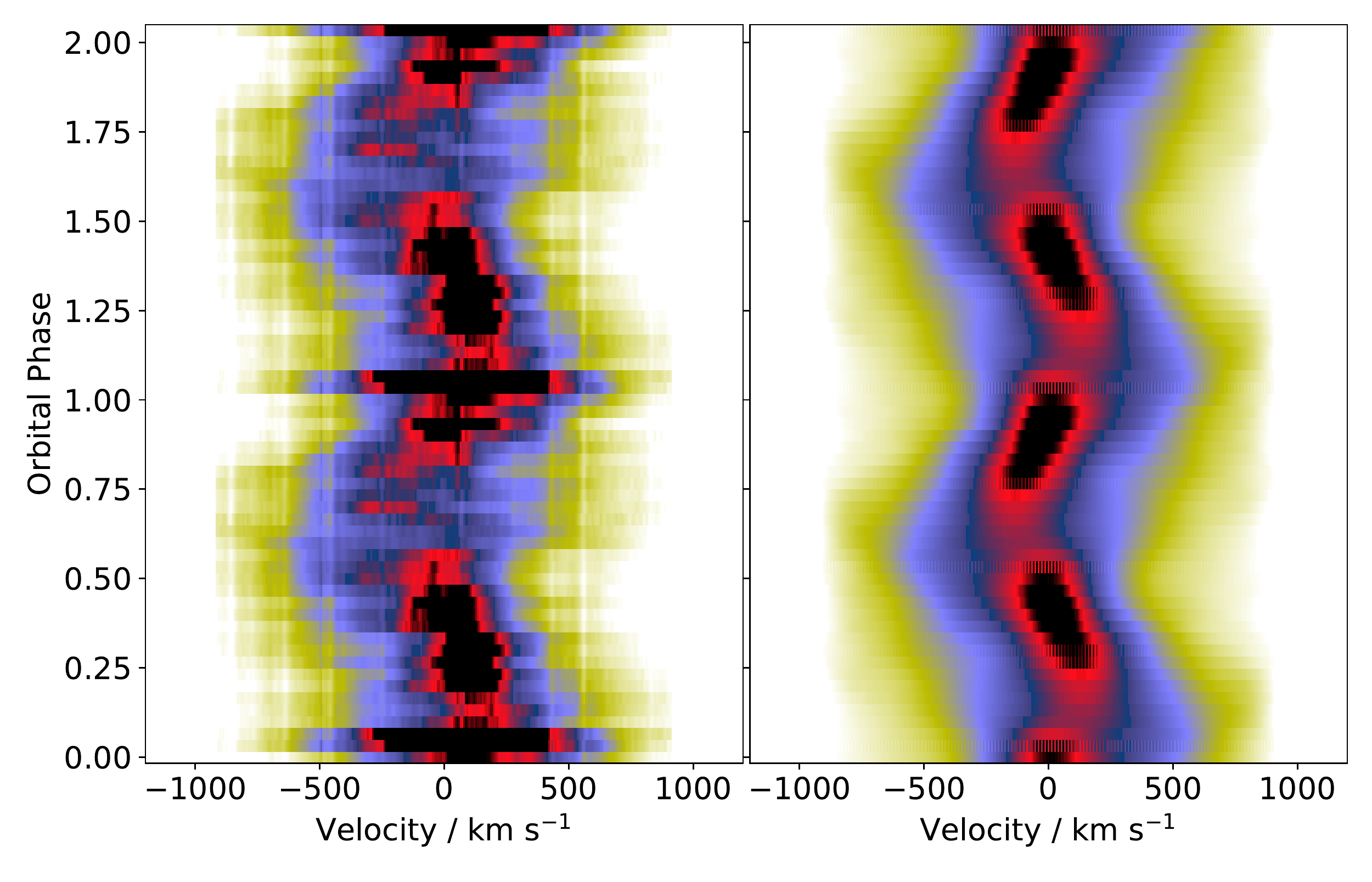}
	\includegraphics[width=1.0\columnwidth]{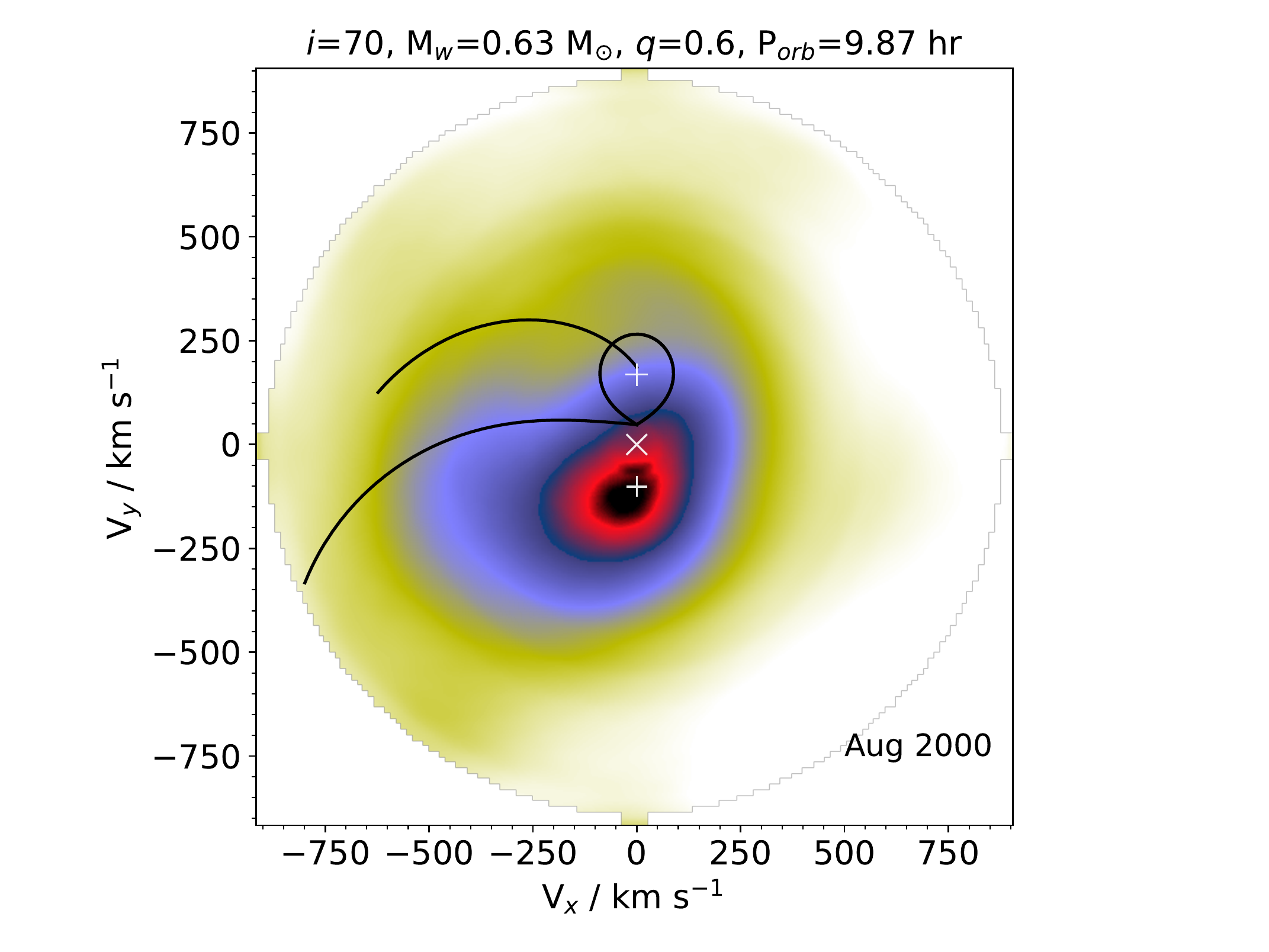}
	
	\includegraphics[angle=0,height=6cm,width=1.0\columnwidth]{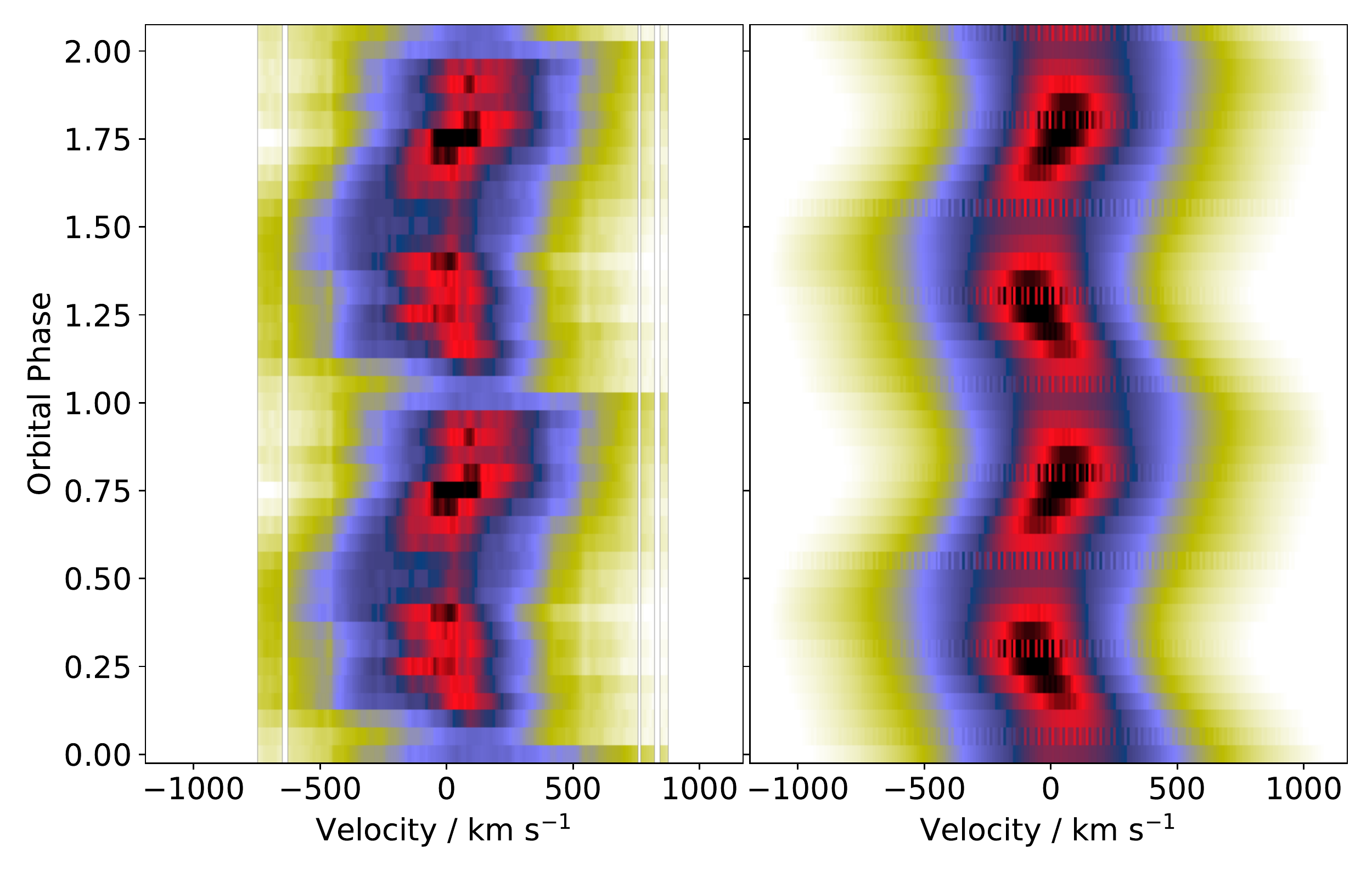}
	\includegraphics[angle=0,width=1.0\columnwidth]{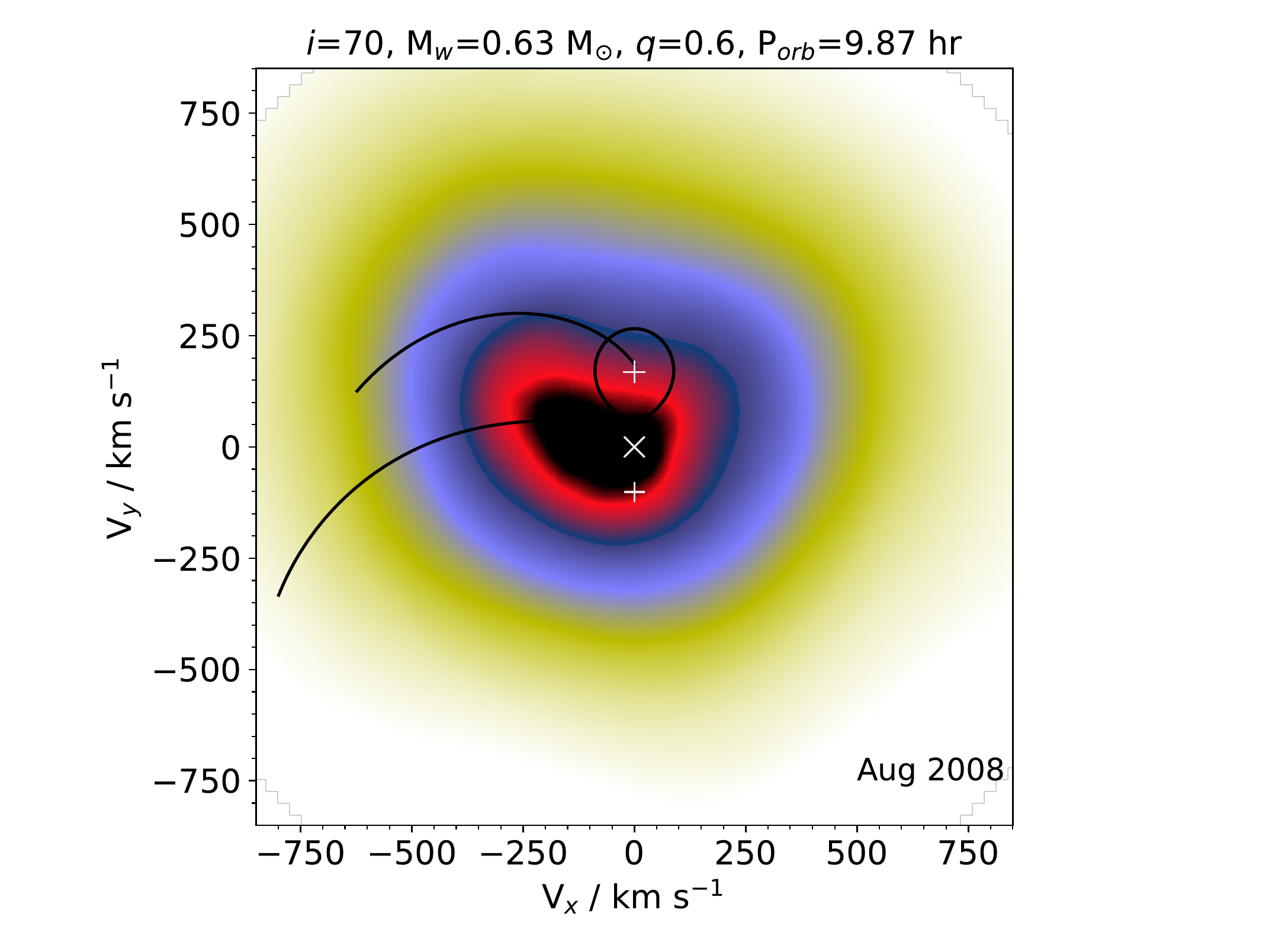}
    \caption{\textit{Top:} Trailed spectrum of the H$\alpha$ emission line for the observed data of 2000 Aug ({\em left}), the reconstructed trailed spectrum ({\em middle}), and the Doppler Tomography ({\em right}). 
    \textit{Bottom:} Trailed spectrum of the H$\alpha$ emission line for the observed data of 2008 Aug({\em left}), the reconstructed trailed spectrum ({\em middle}), and the Doppler Tomography ({\em right}). The relative flux is depicted in a scale of colours, where black represents the highest intensity, followed by red, then blue, and finally yellow. Various features in the tomographies are marked as follows:  the white crosses are the velocities (from top to bottom) of the
secondary star, the centre of mass and the primary star. The Roche lobe of the secondary is shown around its cross. The Keplerian and ballistic trajectories of the gas stream are marked as the upper and lower curves, respectively. See text for further discussion.}
    \label{fig:dopmap-ha}
\end{figure*}

 The Doppler Tomography of Aug 2000 is exhibited in the top right panel of Figure~\ref{fig:dopmap-ha}. Its respective observed and reconstructed spectrograms are shown in the top left panels. With the same layout, in the bottom panels, we present the Aug 2008 results.\par
 
 The trailed spectra of Aug 2000 shows a broad single-peaked structure all throughout the orbit. The structure becomes especially broad in the orbital phase interval of $\sim$0.00--0.05. There is a decrease of the relative flux of the profile in the interval $\sim$0.55--0.75. The line profile presents a clear sinusoidal modulation, as expected from the radial velocity analysis (See Section \ref{sec:radvel}). Its tomography shows a blob-like emission in the lower quadrants centred close to the velocity coordinates of the white dwarf, with its intensity decreasing radially and lacking azimuthal symmetry. \par
 The trailed spectra of Aug 2008 also exhibits a single-peaked structure, with a decrease of relative flux in the intervals spanning $\sim$0.00--0.10 and $\sim$0.40--0.55. The sinusoidal modulation in this run is also evident. The Doppler Tomography again shows an asymmetric blob, with maximum intensity in the lower left quadrant centred at $\sim$(-50,-10), from where the flux decreases radially outward. \par
 The overall structure of both tomograms is very similar to those obtained by \citet{welsh:1998}. These authors interpret such structure, concentrated in the lower quadrants, as a consequence of the propeller action caused by the rapidly-rotating magnetic white dwarf.

\subsection{H{\ensuremath{\beta}} Line Profile}
 \label{subsec:tomo-hb}
\begin{figure*}

	\includegraphics[angle=0,height=18cm,width=16cm]{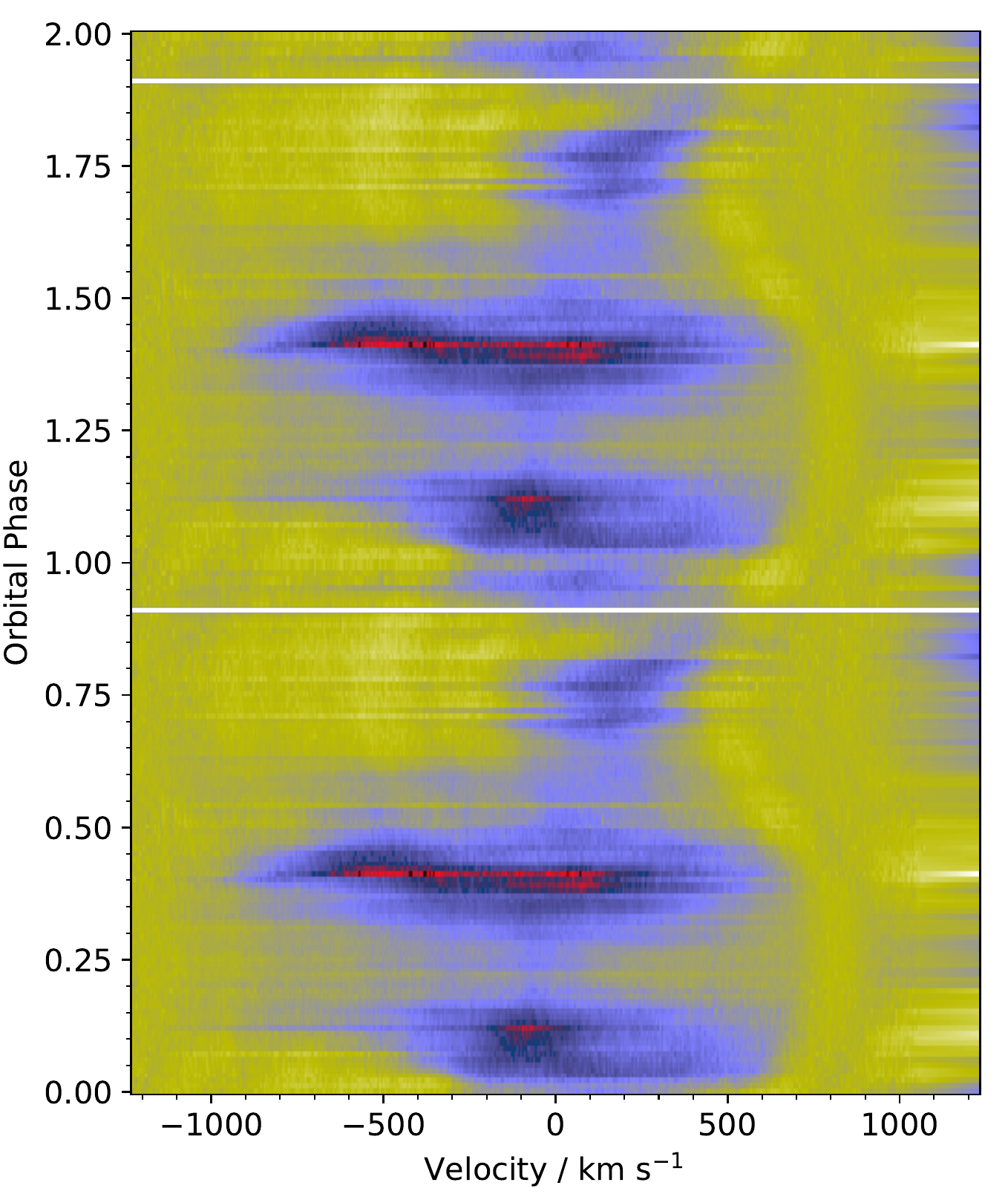}
    \caption{Phase resolved line profiles of H$\beta$.The relative flux is depicted in a scale of colours, where black represents the highest intensity, followed by red, then blue, and finally yellow. For illustrative purposes the orbit is reproduced a second time.}

    \label{fig:profiles-hb}
\end{figure*}

The highly-variable and disrupted nature of the $H\beta$ emission line  profile along the orbital phase, exhibited in the observed trailed spectrum in Figure~\ref{fig:profiles-hb}, did not allow us to produce a dependable tomography. The spectrogram shows a broad single-peaked profile that shifts towards negative velocities in the orbital phase interval of $\sim$0.00-0.20. The profile suddenly narrows at phase $\sim$0.25; but it then considerably broadens in the interval $\sim$0.30-0.50, while drifting towards positive velocities. Within the phases $\sim$0.50-0.90, the single-peaked profile again narrows down, and the midpoint of the emission is abruptly displaced to positive velocities. In this same narrow stage the profile briefly oscillates, initially moving in a negative direction, reaching a minimum velocity at phase $\sim$0.75, and again drifting towards the positive direction until phase $\sim$0.90. At this stage the trail becomes disrupted with the profile \textit{leaping} back to the left and recovering a broader structure, which is preserved until the end of the orbit. \par
This complex and asymmetric behaviour exhibited in the spectrogram not only impeded the construction of a tomogram, but also made  difficult to produce a radial velocity curve, which could explain the highly discrepant semiamplitude value of $K_1~=~185~\pm~11~kms^{-1}$, obtained for this emission line in Section~\ref{sec:radvel-hb}.

\section{Discussion}
\label{discussion}
We made a complementary study to that performed in Paper I, using the original spectroscopic data, by analysing the $H\alpha$ and $H\beta$ Balmer emission lines.
As demonstrated by \citet{welsh:1998}, tracking the orbital motion of the white dwarf using the emission lines is a very unreliable method for AE Aqr. However the propeller model \citep[see][and references therein]{Wynn_1997} does predict a small fraction of accretion of the material onto the white dwarf, meaning that not all of it is ejected by the propeller action.
Also, the narrow component of the Balmer emission lines observed by \citet{RB:1994} is in anti-phase to the absorption lines of the red dwarf, which suggests its origin is near
the primary star. Furthermore, the presence of an accretion disc with a variable nature cannot be discarded, as suggested by the axisymmetric magnetohydrodynamics simulations put forward by \citet{blinova-2019}.\par
With this in mind, in Section~\ref{sec:radvel}, we performed several radial velocity analyses of the emission lines, from which
 our main result was to find a reasonable value of $K_1~=~114~\pm~8~kms^{-1}$ for $H\alpha$, during the observing run of Aug 2000. This value is consistent within $2\sigma$ with the indirect measurements obtained in Paper I and  \citet{eracleous:1994}, whose analyses yielded $K_1=101\pm3~kms^{-1}$ and $K_1=102\pm2~kms^{-1}$, respectively. This consistency implies that, at least for Aug 2000, we were able to track the orbit of the white dwarf by tracing the wings of the emission line profile (which presumably arise from high velocity material orbiting close to the primary star). This can be interpreted as possible observational evidence of material  orbiting the white dwarf.  This claim is further supported by the fact that, during the observations of Aug 2000, the Doppler Tomography shows the emission centred around the position of the primary star. Moreover, the observations of Aug 2008 yielded a value of $K_1=115 \pm 6kms^{-1}$, in good agreement with the 2000 run.  \par
Even if we did detect material orbiting the primary star,  our spectrograms and Tomograms do not show the characteristic signatures of a fully formed accretion disc \citep{Marsh:1988}. In fact, the overall shape of our tomograms was consistent with that exhibited in \citet{welsh:1998}, where the concentration of emission in the lower left quadrant and the azimuthal asymmetry is conjectured to be caused by the propeller mechanism \citep[e.g.][]{Wynn_1997,INB:2004}. Although, as mentioned above, in Aug 2000 we find the centre of emission near the position of the white dwarf; a feature not observed previously nor expected in the propeller model. 
\par
For the observing run of Aug 1991, we made a radial velocity analysis of $H\beta$, obtaining  $K_1=185\pm11ms^{-1}$. This value disagrees with the $H\alpha$ results. However,
we attribute this discrepancy to the highly variable behaviour of $H\beta$ exhibited in its trailed spectrum (see Section \ref{subsec:tomo-hb}).

\section{Conclusions}
Using the spectroscopic data gathered for several epochs in Paper I, we have studied the emission lines of AE~Aqr and  found observational evidence of material orbiting the rapidly rotating white dwarf. Specifically, the radial velocity analysis of the most stable of our observing runs (Aug 2000), yields a semi-amplitude value consistent with previously published results obtained via indirect methods. Also, the Doppler Tomography obtained from this run, shows that the emission is centred around the position of the white dwarf. This is not characteristic of the predictions made in the propeller model.

It is made evident in our analysis of the $H\beta$ emission line, that considerable limitations are posed when tracking the velocity of the emission lines. However, due to the feasibility that these lines are indeed originated in the regions around the primary component, we consider our best estimate of the semi-amplitude value obtained for $H\alpha$, of $K_1~=~114~\pm~8~kms^{-1}$, as a good approximation of the orbital motion of the white dwarf.

\label{conclusions}

\section*{Acknowledgements}

The authors are indebted to DGAPA (Universidad Nacional Aut\'onoma de M\'exico) support, PAPIIT projects IN114917 and IN103120. This research made use of {\sc astropy}, a community-developed core {\sc python} package for Astronomy \citep{Astropy-Collaboration:2013aa}, Python's SciPy signal processing library \citep[][]{virtanen:2020} and {\sc matplotlib} \citep{Hunter:2007aa}. We would also like to thank the anonymous referee, whose comments helped improve the content of this article.

\section*{Data availability}
The data underlying this article can be shared on request to the corresponding author.




\bibliographystyle{mnras}
\bibliography{bibliography.bib} 



\bsp	
\label{lastpage}
\end{document}